\documentclass[11pt]{article}
\usepackage{amsmath}
\usepackage{amsfonts}
\usepackage[english]{babel}
\usepackage{amssymb}
\usepackage{graphicx}
\usepackage[a4paper]{geometry}
\usepackage{hyperref}

\newcommand{\be}{\begin{equation}}
\newcommand{\ee}{\end{equation}}
\newcommand{\bea}{\begin{eqnarray}}
\newcommand{\eea}{\end{eqnarray}}
\newcommand{\vsp}{\vspace{0.4cm}}

\newcommand{\syst}{\mathit{S}}

\newcommand{\stsp}{\mathcal{P}}
\newcommand{\obsp}{\mathfrak{O}}

\title{Time, classical and quantum}

\author{
P. Aniello$^{1,2}$,  F.M. Ciaglia$^{1,2}$,  F. Di Cosmo$^{1,2}$,  G. Marmo$^{1,2}$, J.M. Perez-Pardo$^{2}$\\
\small{} \\
\footnotesize{$^{1}$ Dipartimento di Fisica ``Ettore Pancini'', Universit\`{a} di Napoli ``Federico II''} \\
\footnotesize{Complesso Universitario di Monte S. Angelo, via Cintia, I-80126 Napoli, Italy}\\
\small{}
\footnotesize{$^{2}$ INFN - Sezione di Napoli} \\
\footnotesize{Complesso Universitario di Monte S. Angelo, via Cintia, I-80126 Napoli, Italy}\\
}

\date{}

\begin{document}

\maketitle

\begin{abstract}
We propose a new point of view regarding the problem of time in quantum mechanics, based on the idea of replacing the usual time operator $\mathbf{T}$ with a suitable real-valued function $T$ on the space of physical states.
The proper characterization of the function $T$ relies on a particular relation with the dynamical evolution of the system rather than with the infinitesimal generator of the dynamics (Hamiltonian).
We first consider the case of classical Hamiltonian mechanics, where observables are functions on phase space and the tools of differential geometry can be applied.
The idea is then extended to the case of the unitary evolution of pure states of finite-level quantum systems by means of the geometric formulation of quantum mechanics.
It is found that $T$ is a function on the space of pure states which is not associated to any self-adjoint operator.
The link between $T$ and the dynamical evolution is interpreted as defining a simultaneity relation for the states of the system with respect to the dynamical evolution itself.
It turns out that different dynamical evolutions lead to different notions of simultaneity, i.e., the notion of simultaneity is a dynamical notion.
\end{abstract}

\section{Introduction}

The problem of time in quantum mechanics is a beautiful and subtle one.
We can formalize it with a simple question, namely, is there a self-adjoint operator we can associate to time in quantum mechanics?
Or, even better, is time a quantum observable? 

There are many experimental instances in which this question makes sense because time seems to acquire an observable character.
For example, we can think of the time of arrival of a particle in a detector, the time of occurence of a specific event, or the tunneling time of a particle under the influence of a potential barrier.

In standard quantum mechanics, observable quantities are described by means of self-adjoint linear operators on the Hilbert space of the system.
In this setting, a time observable $\mathbf{T}$ would be characterized as a self-adjoint operator $\mathbf{T}$ which is canonically conjugated to the Hamiltonian operator $\mathbf{H}$ of the system:

\be
\left[\mathbf{H}\,,\mathbf{T}\right]=-\imath\,\hbar\,\mathbb{I}\,.
\ee
In contrast with CCRs relating position and momentum, the commutation relation between $\mathbf{T}$ and $\mathbf{H}$ is plagued by severe technical difficulties.
In \cite{pauli-general_principles_of_quantum_mechanics}, Pauli realized that a self-adjoint operator $\mathbf{T}$ canonically conjugated to the Hamiltonian operator $\mathbf{H}$ does not exist whenever the spectrum of $\mathbf{H}$ is bounded from below.
Pauli's proof was not rigorous, and, to be fair, he never claimed it to be so.
However, it took some time for the rigorous mathematical formulation of the problem to be settled (see \cite{galapon-pauli's_theorem_and_quantum_canonical_pairs}, and \cite{srinivas_vijayalakshmi-the_time_of_occurrence_in_quantum_mechanics}), and, in the meantime, different strategies to cope with the problem have been proposed.
For instance, attention has been given to the possibility of relaxing the self-adjointness condition for the time observable $\mathbf{T}$.
In this direction, of particular interest is the construction of a maximally symmetric time operator $\mathbf{T}$ which is canonically conjugated to the Hamiltonian operator $\mathbf{H}$ of the $1$-dimensional free particle given by Ahronov and Bohm in \cite{aharonov_bohm-time_in_the_quantum_theory_and_the_uncertainty_relation_for_time_and_energy}.
This operator is a sort of canonical quantization of the classical passage time of Newtonian mechanics, and thus, its physical interpretation is related to the experimental concepts of passage time, and of time of flight.
Another change of perspective occured, and efforts were, and are made to construct a positive operator-valued measure (POVM) having a particular covariance property with respect to the dynamics, and that can be reasonably interpreted as a time POVM (\cite{brunetti_fredenhagen_hoge-time_in_quantum_physics_from_an_external_parameter_to_an_intrinsic_observable}, \cite{busch_grabowski_lahti-time_observables_in_quantum_theory}, \cite{a.a.v.v.-time_in_quantum_mechanics}, and \cite{a.a.v.v.-time_in_quantum_mechanics_volume_2}).
In this setting, the physical interpretation of the time POVM constructed in \cite{brunetti_fredenhagen-time_of_occurrence_observable_in_quantum_mechanics} is related to the experimental concept of time of occurrence.
Finally, some interesting counterexamples to Pauli's theorem have been given.
Among the most interesting ones  is the case of a phase operator constructed by Galindo \cite{galindo-phase_and_number} and Garrison and Wong \cite{garrison_wong-canonically_conjugate_pairs_uncertainty_relations_and_phase_operators}, which is a bounded, self-adjoint operator canonically conjugated to the number operator, and thus with the Hamiltonian operator, of the $1$-dimensional quantum harmonic oscillator.
The physical interpretation of this operator is in some sense related to the quantum-mechanical formulation of the action-angle variables exposed by Dirac in \cite{dirac-the_elimination_of_the_nodes_in_quantum_mechanics}.

From this brief discussion we can extract two important facts.
First of all, time in quantum mechanics is a dynamical quantity which is intimately connected with the specific dynamical evolution of the system and with specific experimental questions.
Second, it seems that self-adjoint operators are simply not enough to handle the problem of time in quantum mechanics, and different mathematical objects may be appropriate to treat different aspects of time.
In this article, we focus on the simultaneity aspect of time in quantum mechanics, and  propose to describe it by means of a real-valued function $T$ on the space of physical states, which we call a time function, satisfying a particular equivariance condition with respect to the dynamical evolution of the system.

\vsp

In accordance with Einstein's theory of special relativity, we recognize two different but related aspects of our common perception of time in physical phenomena. 
On the one hand, time appears as an evolution parameter, a sort of ordering label by means of which we formalize the perception of the causal aspect of ``before'' and ``after''.
Following an heuristic argument, the mathematical object that captures this aspect of time in a spacetime framework is a vector field, say $\frac{\partial}{\partial t}$.
Given an integral curve $\gamma_{m}(\tau)$ of $\frac{\partial}{\partial t}$ starting at $m=\gamma_{m}(0)$, the parameter $\tau$ ``measures'' causality in the sense that $m_{1}=\gamma_{m}(\tau_{1})$ casually precedes $m_{2}=\gamma_{m}(\tau_{2})$ if and only if $\tau_{1}<\tau_{2}$, and thus, all the events lying on $\gamma_{m}(\tau)$  are interpreted as causally connected through the spacetime evolution determined by $\frac{\partial}{\partial t}$.
It is clear that this causal aspect of time is meaningful only in relation to events lying on the same integral curve $\gamma_{m}(\tau)$ of $\frac{\partial}{\partial t}$.

On the other hand, time is naturally associated to the concept of simultaneity, which is a particular relation between different events that need not to lie on the same integral curve of $\frac{\partial}{\partial t}$.
The purpose of simultaneity is to provide a way to compare the evolution through $\frac{\partial}{\partial t}$ of different initial events so that a relational notion of ``before'' and ``after'' is meaningful. 
Consequently, the simultaneity aspect of time can not be described by means of the vector field $\frac{\partial}{\partial t}$, and the correct mathematical object is an integrable differential one-form, say $dt$.
The integrability condition implies that $dt$ defines a codimension-one foliation $\mathcal{F}$ of the spacetime $\mathcal{M}$.
Of course, $dt$ can not be completely arbitrary since simultaneity must take into account the specific spacetime evolution described by $\frac{\partial}{\partial t}$.
Specifically, $dt$ must satisfy:

\be
dt\left(\frac{\partial}{\partial t}\right)=1\,.
\ee
This condition implies that the leaves of the foliation $\mathcal{F}$ induced by $dt$ are transversal to the integral curves of $\frac{\partial }{\partial t}$, and it is precisely this transversality condition that motivates the interpretation of the events on the same leaf as simultaneous events.
For an extensive and rigorous treatment of this notion of simultaneity in a spacetime framework we refer to \cite{deritis_marmo_preziosi-a_new_look_at_relativity_transformations} and \cite{marmo_preziosi-the_structure_of_spacetime_relativity_groups}.

The causality and simultaneity aspects of time encoded in the couple $\left(\frac{\partial}{\partial t}\,,dt\right)$ are purely kinematical, as they are defined with respect to a fixed spacetime background.
This means that we have to lift these considerations to a dynamical setting in order to make contact with quantum mechanics, where time seems to acquire a purely dynamical flavour, and spacetime does not enter directly in the formulation of the theory.

In quantum mechanics, as well as in other dynamical theories, we have two objects that we can use to set up a dynamical framework for simultaneity.
There is the space of states $\stsp$ of the system and its dynamical evolution $\{\phi_{\tau}\}$.
The space of states $\stsp$ plays a role analogous to that of spacetime $\mathcal{M}$ in our previous discussion, while the trajectories of the dynamical evolution $\{\phi_{\tau}\}$ represent the causality aspect of time in analogy with the integral curves of the vector field $\frac{\partial}{\partial t}$.
What is missing is a mathematical object describing simultaneity with respect to $\{\phi_{\tau}\}$, and we propose to identify it with a function $T$ defined on a suitable subset $\stsp_{*}$ of the space of states $\stsp$.
In analogy with the one-form $dt$, the time function $T$ satisfies a suitable equivariance condition with respect to $\{\phi_{\tau}\}$ which resembles the transversality condition between $dt$ and $\frac{\partial}{\partial t}$.
Specifically, $T$ is such that its level sets are mapped into each other by $\{\phi_{\tau}\}$, and thus, all the states in a level set of $T$ are interpreted as simultaneous states.

\vsp

Following this line of thought, we will analyze the case of unitary evolution of the pure states of a finite-level quantum system by means of the so-called geometric formulation of quantum mechanics, according to which, the tools of differential geometry characteristic of classical mechanics can be used in the quantum setting.
Consequently, in Section \ref{section: Time function and simultaneity for classical systems} the idea of a time function $T$ is presented in the classical setting.
A precise definition for the equivariance condition of $T$ with respect to the dynamical evolution is given, and different examples of time function for well-known physical systems are presented.

In Section \ref{section: quantum case} we briefly recall the main points of the geometric formulation of quantum mechanics, and then pass to analyze finite-dimensional systems providing the construction of a family of time functions for all the unitary evolutions generated by a Hamiltonian operator $\mathbf{H}$ with at least two different eigenvalues. 
This is interesting because the construction of a time operator $\mathbf{T}$ canonically conjugated to a Hamiltonian operator $\mathbf{H}$ for a finite-level quantum system is forbidden because of dimensional reasons.
We will find that the time function $T$ is not the expectation value function of some linear operator on the Hilbert space $\mathcal{H}$ of the system.
Hence, $T$ is not associated to an observable in the canonical sense, i.e., a self-adjoint linear operator on $\mathcal{H}$, rather, it is more like other functions on the space of states, such as Entropy or Purity.
We want to stress that the time functions for the unitary evolutions of finite-level quantum systems are not the quantization of some time function defined for a classical system.
Indeed, the geometrical definition of time function we give in Section \ref{section: Time function and simultaneity for classical systems} is general enough to encompass all those physical system described using the tools of differential geometry.

\section{Time function and simultaneity for classical systems} \label{section: Time function and simultaneity for classical systems}

In the description of a classical system $\syst$, the space of states is a finite-dimensional differential manifold $\stsp$, while the algebra of observables $\obsp$ of $\syst$ is realized as an algebra of functions on $\stsp$.
Essentially, we will take $\obsp$ to be the algebra $C^{\infty}(\stsp)$ of real-valued, smooth functions on $\stsp$ endowed with the pointwise product.
The expectation value of an observable $f\in\obsp$ on the state $p\in\stsp$ is just the evaluation $f(p)$ of the function $f$ on the state $p$.
The dynamical evolution of the system is described by the one-parameter group $\{\phi_{\tau}\}_{\tau\in\mathbb{R}}$ generated by a complete vector field $\Gamma$ referred to as the dynamical vector field, and the image $\phi_{\tau}(p)\subset\stsp$ of $p\in\stsp$ through the dynamical evolution is the dynamical trajectory of the initial state $p$.
In the Hamiltonian formulation $\stsp$ is endowed with a Poisson tensor $\Lambda$, and the dynamical vector field $\Gamma$ is the Hamiltonian vector field associated to a Hamiltonian function $H$:

\be
\Gamma=\Lambda(\mathrm{d}H)\,.
\ee
We can divide dynamical evolutions into three different classes according to the nature of their dynamical trajectories.
First of all there are periodic dynamical evolutions, for which all the dynamical trajectories are periodic.
Then, there are non-periodic dynamical evolutions, for which none of the dynamical trajectories are periodic, and finally, there are mixed dynamical evolutions, for which some dynamical trajectories are periodic and some are not.

The time function we want to describe explicitely depends on the dynamical evolution of the system, and, in general, different dynamics will lead to different time functions.
We will now introduce two different types of time function and associated simultaneity relations, one which is well-suited for non-periodic dynamical evolutions, and one which is well-suited for periodic dynamical evolutions (periodic time function).
In both cases, we need to introduce a reduced space of states $\stsp_{*}$, i.e., an open dense subset of $\stsp$ which is invariant with respect to the dynamical evolution and which will be the domain of definition of the time function $T$.
The fact that $\stsp_{*}$ is in general different from the whole space of states $\stsp$ is related to the existence of fixed points for the dynamical evolution $\{\phi_{\tau}\}$ in consideration, i.e., states that are completely unaffected by $\{\phi_{\tau}\}$.
As we will see in Section \ref{section: quantum case}, the case of mixed dynamical evolution can be handled using a periodic time function.

\vsp

In the case of a non-periodic dynamical evolution $\{\phi_{\tau}\}$, we define a function $T:\stsp_{*}\rightarrow\mathbb{R}$ to be a time function if the following conditions are satisfied:

\be\label{eqn: abstract characterization of Time function 1}
T(p)\neq T\left(\phi_{\tau}(p)\right)\;\;\;\;\;\;\forall \tau\neq0\,,
\ee
\be\label{eqn: abstract characterization of Time function}
T(p_{1})=T(p_{2})\;\;\;\Longrightarrow\;\;T\left(\phi_{\tau}(p_{1})\right)=T\left(\phi_{\tau}(p_{2})\right)\,.
\ee
The function $T$ naturally induces an equivalence relation $\sim_{T}$ on $\stsp_{*}$ given by:

\be
p_{1}\sim p_{2}\;\;\;\;\;\mbox{ iff }\;\;\;\;\;T(p_{1})=T(p_{2})\,.
\ee
An equivalence class of $\sim_{T}$ is denoted as $\mathcal{F}_{t}$, with $t\in\mathbb{R}$.
As a set, $\mathcal{F}_{t}$ is given by:

\be
\mathcal{F}_{t}:=\left\{p\in\stsp\,:\;\;T(p)=t\right\}\,.
\ee
Equation \ref{eqn: abstract characterization of Time function} implies that the dynamical evolution of the system is perfectly transversal to the equivalence relation $\sim_{T}$, that is, it ``moves'' the states in the equivalence class $\mathcal{F}_{t}$ into an equivalence class $\mathcal{F}_{t'}$ which is different from the initial one because of equation \ref{eqn: abstract characterization of Time function 1}.
Accordingly, $\sim_{T}$ is interpreted as a simultaneity relation relative to the dynamical evolution $\{\phi_{\tau}\}$, and the states in $\mathcal{F}_{t}$ are interpreted as the simultaneous states defined by $\sim_{T}$.

Note that a similar approach appears in \cite{esposito_marmo_sudarshan-from_classical_to_quantum_mechanics}, where a classical dynamical time is defined as a function $T$ on the phase-space of the system such that $\left\{T\,,H\right\}=\mathcal{L}_{\Gamma}T=1$, where $\{\,,\}$ denotes the Poisson brackets, $H$ is the Hamiltonian function of the system, and $\Gamma$ the dynamical vector field associated to $H$.
Furthermore, the idea of dynamical time, both in classical and quantum theory, is analyzed in \cite{bhamathi_sudarshan-time_as_a_dynamical_variable}.

For the sake of simplicity, here we will only give the explicit form for the time functions of some simple systems without entering into a discussion of the explicit construction of such functions.
However, the detailed construction we will give in Section \ref{section: quantum case}, which is based on the action-angle variables formulation of the dynamics of the system, can easily be adapted to the examples presented here.

Let us consider a point particle in a constant force field.
The space of states of the system is $\stsp=T^{*}\mathbb{R}^{3}\cong\mathbb{R}^{6}$ with global Cartesian coordinates $(\mathbf{q}\,,\mathbf{p})$, and the dynamical evolution $\{\phi_{\tau}\}$ is generated by the complete vector field:

\be
\Gamma=\sum_{j=1}^{3}\,\left(\frac{p_{j}}{m}\frac{\partial}{\partial q_{j}} + F_{j}\frac{\partial}{\partial p_{j}}\right)\,.
\ee
It is clear that $\{\phi_{\tau}\}$ has no fixed points since $\Gamma$ has no zeros.
The expressions of the dynamical trajectories of the system in the coordinates system $(\mathbf{q}\,,\mathbf{p})$ read:

\be
\phi_{\tau}\left(q_{j}\,, p_{j}\right)=\left(\frac{F_{j}}{2m}\tau^{2} + \frac{p_{j}}{m}\tau + q_{j}\,, F_{j}\tau + p_{j}\right)\;\;\;\mbox{ with } j=1,2,3\,.
\ee
Let $T:\stsp\rightarrow\mathbb{R}$ be the function:

\be
T(\mathbf{q}\,,\mathbf{p})=\frac{\mathbf{F}\cdot\mathbf{p}}{F^{2}}\,.
\ee
An explicit calculation shows that:

\be
T\left(\phi_{\tau}(\mathbf{q}\,,\mathbf{p})\right)=\tau + \frac{\mathbf{F}\cdot\mathbf{p}}{F^{2}}\,,
\ee
which means that $T$ is a time function for the dynamical evolution considered.

Now, let us consider the free point particle.
The space of states of the system is $\stsp=T^{*}\mathbb{R}^{3}\cong\mathbb{R}^{6}$ with global Cartesian coordinates $(\mathbf{q}\,,\mathbf{p})$, and the dynamical evolution is generated by the complete vector field:

\be
\Gamma=\sum_{j=1}^{3}p_{j}\frac{\partial}{\partial q_{j}}\,.
\ee
The explicit form of the dynamical trajectories of the system in the coordinates system $(\mathbf{q}\,,\mathbf{p})$ reads:

\be
\phi_{\tau}\left(q_{j}\,, p_{j}\right)= \left(p_{j}\tau + q_{j}\,, p_{j}\right)\;\;\;\mbox{ with } j=1,2\,.
\ee
In this case, the dynamical evolution presents fixed points since $\Gamma$ has zeros.
Specifically, every state $(\mathbf{q}\,,\mathbf{0})$ is a fixed point of $\{\phi_{\tau}\}$.

Consequently, we have to define the reduced space of states $\stsp_{*}$ as the space of states $\stsp$ without the fixed points:

\be
\stsp_{*}:=\left\{(\mathbf{q}\,,\mathbf{p})\in\stsp\colon (\mathbf{q}\,,\mathbf{p})\neq (\mathbf{q}\,,\mathbf{0})\right\}\,.
\ee 
In this case, a possible time function for the system is given by the time of arrival of Newtonian mechanics:

\be
T(\mathbf{q}\,,\mathbf{p})=\frac{\mathbf{p}\cdot\mathbf{q}}{p^{2}}\;\;\;\Longrightarrow\;\;\; T\left(\phi_{\tau}(\mathbf{q}\,,\mathbf{p})\right)=\tau + \frac{\mathbf{p}\cdot\mathbf{q}}{p^{2}}\,.
\ee
Note that this result is in accordance with \cite{esposito_marmo_sudarshan-from_classical_to_quantum_mechanics}.

\vsp

In the case of a periodic dynamical evolution $\{\phi_{\tau}\}$, we need to choose a different target space for the function $T$ in order to handle the periodic trajectories of the system.
Specifically, we chose the target space to be the one-dimensional torus $\mathbb{T}$, and define a function $T:\stsp_{*}\rightarrow \mathbb{T}$ to be a periodic time function $T$ with period $\tau_{T}$ if the following conditions are satisfied: 

\be
T\left(\phi_{\tau_{1}}(p)\right)=T\left(\phi_{\tau_{2}}(p)\right)\;\;\;\;\;\;\mbox{ iff } \;\;\; \tau_{2}=\tau_{1} + k\tau_{T}\,,\;k\in\mathbb{Z} \,,
\ee
\be
T(p_{1})=T(p_{2})\;\;\;\Longrightarrow\;\;T\left(\phi_{\tau}(p_{1})\right)=T\left(\phi_{\tau}(p_{2})\right)\,.
\ee
Because of the global non-trivial topology of the torus $\mathbb{T}$, the simultaneity relation $\sim_{T}$ associated to $T$ becomes periodic, and the closed trajectories of the system can be handled accordingly.
Furthermore, the periodicity of $T$ needs not to be the periodicity of $\{\phi_{\tau}\}$, and thus we can manage systems admitting periodic trajectories with different periods using the same periodic time function.

The paradigmatic system for which a periodic time function is needed is the one-dimensional harmonic oscillator on $\stsp=T^{*}\mathbb{R}\cong\mathbb{R}^{2}$.
In Cartesian coordinates $(q\,,p)$ the dynamical evolution is generated by the complete vector field $\Gamma$:

\be
\Gamma=\frac{p}{m}\frac{\partial}{\partial q} - m\nu^{2}q\frac{\partial}{\partial p}\,.
\ee
The dynamical trajectories of the system are:

\be
\phi_{\tau}\left(q\,, p\right)=\left(q\cos(\nu\tau) - \frac{p}{m\nu}\sin(\nu\tau)\,, - m\nu q\sin(\nu\tau) - p\cos(\nu\tau)\right)\,,
\ee
and the only fixed point of the dynamical evolution is the origin $(0\,,0)$ itself.

We define the reduced space of states as $\stsp_{*}:=\stsp-\{(0\,,0)\}$, and note that there is a natural diffeomorphism:

\be
\Psi:\stsp_{*}\longrightarrow \mathbb{T}\times\mathbb{R}^{+}\,.
\ee
Using a local coordinates system $(\vartheta\,,H)$ on $\mathbb{T}\times\mathbb{R}^{+}$, the local expression of $\Psi$ reads:

\be
(q\,,p)\mapsto\Psi(q\,,p)=\left(\vartheta=\arctan\left(\nu m\frac{q}{p}\right)\,,H=\frac{p^{2}}{2m} + \frac{m\nu^{2}q^{2}}{2}\right)\,.
\ee
A direct calculation shows that the local expression of the dynamical vector field $\widetilde{\Gamma}=\Psi_{*}\Gamma$ on $\mathbb{T}\times\mathbb{R}^{+}$ with respect to $(\vartheta\,,H)$ is:

\be
\widetilde{\Gamma}=\nu\frac{\partial}{\partial\vartheta}\,,
\ee
and thus the local expression of the dynamical trajectories is:

\be
\Psi\left(\phi_{\tau}(q\,, p)\right)=\left( \nu\tau + \vartheta_{0}\,, H_{0}\right)\,.
\ee
It is then clear that the function $T:\stsp_{*}\rightarrow \mathbb{T}$ defined as:

\be
T:=pr^{\mathbb{T}}\circ\Psi\,,
\ee
is a periodic simultaneity function for $\{\phi_{\tau}\}$ with period $\tau_{T}=\frac{\nu}{2\pi}$.
Note that $T$ is a submersion.
The simultaneous states associated to $\sim_{T}$ can be described by means of a particular one-form $\Theta$ on $\stsp$.
To define $\Theta$, denote with $\theta$ the differential one-form on the torus $\mathbb{T}$ which is dual to the globally defined vector field on $\mathbb{T}$ generating the action of the torus on itself.
The one-form $\theta$ is a closed but not exact one-form.
Then, define $\Theta$ as the pullback of $\theta$ by means of $T$, i.e.:

\be
\Theta:=T^{*}\theta=\frac{1}{H}\left(pdq - qdp\right)\,.
\ee
This is a closed but not exact one-form on $\stsp_{*}$, hence, it gives rise to a foliation $\mathcal{F}$ of $\stsp_{*}$, and the leaves of this foliation are precisely the simultaneous states defined by $T$.
In this case, the leaves are just the radial lines in $\stsp_{*}$ approaching the origin $(0\,,0)$ without ever reaching it.

\section{Time function for finite-level quantum systems} \label{section: quantum case}

We will now extend the ideas exposed in the previous section to the finite-dimensional quantum case.
Our approach is based on the so-called geometric formulation of quantum mechanics (\cite{ashtekar_schilling-geometrical_formulation_of_quantum_mechanics}, \cite{ercolessi_marmo_morandi-from_the_equations_of_motion_to_the_canonical_commutation_relations}, \cite{grabowski_kus_marmo-geometry_of_quantum_systems_density_states_and_entanglement}, \cite{carinena_ibort_marmo_morandi-geometry_from_dynamics_classical_and_quantum}), according to which the mathematical methods of differential geometry characteristic of classical mechanics, can be used in the quantum context.

The space of pure states of a finite-level quantum system with Hilbert space $\mathcal{H}\cong\mathbb{C}^{n}$ is the complex projective space $P(\mathcal{H})$ associated to $\mathcal{H}$.
An element $p_{\psi}\in P(\mathcal{H})$ is an equivalence class of non-null vectors in $\mathcal{H}$ with respect to the equivalence relation:

\be
|\psi\rangle\sim|\varphi\rangle\;\;\;\mbox{ iff }\;\;\; |\psi\rangle=\alpha|\varphi\rangle\,,\;\;\alpha\in\mathbb{C}_{0}\,.
\ee
The set $P(\mathcal{H})$ is endowed with the quotient topology, and we denote by $\pi$ the continuous projection from $\mathcal{H}_{0}$ onto $P(\mathcal{H})$.

From a geometrical point of view, $P(\mathcal{H})$ is a real, $2(n-1)$-dimensional Kaehler manifold.
The Kaehler structure of $P(\mathcal{H})$ is encoded in three geometrical objects, namely, a symplectic structure $\omega$, a Riemannian metric $g$, and a complex structure $J$ satisfying the compatibility condition:

\be
g\left(X\,,Y\right)=\omega\left(J(X)\,,Y\right)\;\;\;\;\;\;\forall X,Y\in\mathfrak{X}\left(P(\mathcal{H})\right)\,.
\ee

In this framework, to every self-adjoint operator $\mathbf{A}\in\mathcal{B}(\mathcal{H})$ there is associated a real-valued function $e_{\mathbf{A}}:P(\mathcal{H})\rightarrow\mathbb{R}$ by means of:

\be
e_{\mathbf{A}}(p_{\psi}):=\frac{\langle\psi|\mathbf{A}|\psi\rangle}{\langle\psi|\psi\rangle}\,.
\ee
Accordingly, $e_{\mathbf{A}}$ is nothing but the expectation value of $\mathbf{A}$ on the state $p_{\psi}$.
In this way, observables of quantum mechanics are represented by means of expectation value functions on the complex projective space.
The symplectic structure $\omega$ and the Riemannian metric $g$ allow for the definition of two vector fields naturally associated to $e_{\mathbf{A}}$, namely:

\be
X_{\mathbf{A}}:=i_{\mathrm{d}e_{\mathbf{A}}}\Lambda\;\;\;\;\;\;\;\;\;\;\;\;Y_{\mathbf{A}}:=i_{\mathrm{d}e_{\mathbf{A}}}G\,,
\ee
where $\Lambda=\omega^{-1}$ and $G=g^{-1}$.
Furthermore, the tensor $\Lambda$ allows to define of a Poisson bracket $\{\,,\}$ on the algebra of smooth functions.
On the expectation value functions the bracket reads:

\be
\{e_{\mathbf{A}}\,,e_{\mathbf{B}}\}=\Lambda\left(\mathrm{d}e_{\mathbf{A}}\,,\mathrm{d}e_{\mathbf{B}}\right)=e_{\imath\left[\mathbf{A}\,,\mathbf{B}\right]}\,,
\ee
where $\left[\,,\right]$ denotes the commutator of linear operators.

The action $|\psi\rangle\mapsto\mathbf{U}\,|\psi\rangle$ of the unitary group $\mathcal{U}(\mathcal{H})\cong\mathcal{U}(n)$ on $\mathcal{H}$ induces the action $p_{\psi}\mapsto \widetilde{p_{\psi}}=p_{\mathbf{U}\psi}$ of $\mathcal{U}(\mathcal{H})$ on $P(\mathcal{H})$, and it turns out that the fundamental vector fields of this action are precisely the Hamiltonian vector fields $X_{\mathbf{A}}$ associated to each $e_{\mathbf{A}}$, with $\mathbf{A}\in\mathcal{B}(\mathcal{H})$ and such that $\mathbf{U}=\exp(-\imath\mathbf{A})$.
Accordingly, the dynamical evolution generated by the self-adjoint Hamiltonian operator $\mathbf{H}$ is written in geometrical language as the one-parameter group of diffeomorphisms $\{\phi_{\tau}\}$ of $P(\mathcal{H})$ generated by this dynamical vector field:

\be
\Gamma\equiv X_{\mathbf{H}}:=i_{\mathrm{d}e_{\mathbf{H}}}\Lambda\,.
\ee
From the physical point of view, this dynamical evolution describes a closed quantum system.

In general, the unitary evolution can be periodic or mixed, and we will show that every such dynamical evolution admits a family of periodic simultaneity functions.
Each of which is given by a submersion $T:\stsp_{*}\rightarrow \mathbb{T}$, where $\stsp_{*}$ is an open submanifold of $\stsp=P(\mathcal{H})$ and is invariant with respect to the dynamical evolution.
This is intimately connected with the fact that the dynamical system is integrable in the sense of Liouville-Arnol'd, i.e., it always admits $(n-1)$ functionally independent constants of the motion in involution, and it is always possible to find an open submanifold $\stsp_{*}$ of $\stsp$ which is invariant for the dynamical evolution, and for which a formulation in terms of action-angle variables is possible.
The family of periodic simultaneity functions is related to the submersions arising from the projection onto one of the $(n-1)$ Liouville's tori of the system.

\vsp

Let $\mathbf{H}$ denote the Hamiltonian operator of the system.
Suppose its spectrum $\sigma(\mathbf{H})$ contains at least two different eigenvalues, and construct an orthonormal basis $\{|j\rangle\}_{j=1,...,n}$ on $\mathcal{H}$ consisting of eigenvectors of $\mathbf{H}$.
Write $\mathbf{H}=\sum_{j}\nu_{j}\mathbf{E}_{j}$, with $\nu_{j}$ the eigenvalues of $\mathbf{H}$, and $\mathbf{E}_{j}$ the projector onto the subspace in $\mathcal{H}$ spanned by the $j$-th eigenvector.
The operator $\mathbf{E}_{j}$ is self-adjoint for all $j$, and $\left[\mathbf{E}_{j}\,,\mathbf{E}_{k}\right]=\mathbf{0}$ for all $j$ and $k$.
Consequently, the functions $e_{\mathbf{E}_{j}}\equiv e_{j}$ are constants of the motion in involution, i.e., $\{e_{j}\,,H\}=0$ for all $j$, and $\{e_{j}\,,e_{k}\}=0$ for all $j$ and $k$.
Moreover, the vector fields $X_{j}:=X_{\mathbf{E}_{j}}$ are pairwise commuting, and the dynamical vector field $\Gamma$ reads:

\be
\Gamma=\sum_{j}\nu_{j}X_{j}\,.
\ee
The vector fields $X_{j}$ are complete and, since their flows are periodic, each of them separately defines an action of a torus $\mathbb{T}$ on $\stsp$.
However, they are not all independent because $\sum_{j}\mathbf{E}_{j}=\mathbb{I}$, which means that $\sum_{j}X_{j}=0$, and the same holds true for the functions $e_{j}$, that is, $\sum_{j}e_{j}=1$.
Nevertheless, we can always find $(n-1)$ independent vector fields and functions among them.

Let us choose the first $(n-1)$ vector fields and the first $(n-1)$ functions, that is, set:

\be
X_{n}=-\sum_{j=1}^{n-1}\,X_{j}\;\;\;\;\;\;e_{n}=1-\sum_{j=1}^{n-1}\,e_{j}\,,
\ee
and write the dynamical vector field as:

\be
\Gamma=\sum_{j=1}^{n-1}\,(\nu_{j}-\nu_{n})X_{j}\,.
\ee
Using the $(n-1)$ constants of the motion we will now show that there is an open submanifold $\stsp_{*}\subset\stsp$ for which a global action-angle variables formulation of the problem is possible.
At this purpose, note that the critical points of $e_{j}$, that is, the points $p_{\psi}\in\stsp$ for which $\mathrm{d}e_{j}(p_{\psi})=0$, are just the zeros of the vector field $X_{j}$, and these are all those states $p_{\psi}$ for which $\langle j|\psi\rangle=0$ or $\langle k|\psi\rangle=0$ for all $k\neq j$, where $|j\rangle$ and $|k\rangle$ denote, respectively, the $j$-th and $k$-th normalized eigenvector of $\mathbf{H}$.
From this it follows that the set:

\be
\stsp_{*n}:=\left\{p_{\psi}\in\stsp:\;\mathrm{d}e_{1}(p_{\psi})\wedge \mathrm{d}e_{2}(p_{\psi})\wedge\cdots\wedge \mathrm{d}e_{n-1}(p_{\psi})\neq0\right\}
\ee
consists of all those states $p_{\psi}$ such that $\langle j|\psi\rangle\neq0$ for all $j$, that is, the vector $\psi$ of which $p_{\psi}$ is the associated ray, must have non-zero components with respect to every normalized eigenvector of $\mathbf{H}$.
Had we started with a different choice of $(n-1)$ functions and vector fields, say, $X_{k}=-\sum_{j\neq k}\,X_{j}$ and $e_{k}=1-\sum_{j\neq k}\,e_{j}$, the set $\stsp_{*k}$ would have been the same as $\stsp_{*n}$, therefore, the explicit choice of $(n-1)$ independent functions and vector fields is irrelevant, and we will simply write:

\be
\stsp_{*}:=\left\{p_{\psi}\in\stsp:\;\;\langle j|\psi\rangle\neq0\;\forall j=1,...n\right\}\,.
\ee
The set $\stsp_{*}$ is an open subset of $\stsp$, and thus an open submanifold of $\stsp$ on which there are $(n-1)$ linearly independent constants of the motion in involution.
Let $F:\stsp\rightarrow\mathbb{R}^{n-1}$ be given by $p_{\psi}\mapsto\left(e_{1}(p_{\psi})\,,\cdots\,,e_{n-1}(p_{\psi})\right)$.
Every $\mathbf{a}=F(p_{\psi})$ with $p_{\psi}\in\stsp_{*}$ is a regular value, hence, $F^{-1}(\mathbf{a})$ is a closed submanifold of $\stsp$.

Since $F^{-1}(\mathbf{a})$ is a closed subset of $\stsp$, and $\stsp$ is a compact manifold, we have that $F^{-1}(\mathbf{a})$ is a compact submanifold of $\stsp$.
Then, according to the Liouville-Arnold's theorem, we have the diffeomorphism:

\be
\Psi:\stsp_{*}\longrightarrow\left(\bigcup_{\alpha} \mathbb{T}^{n-1}_{\alpha}\right)\times I^{n-1}\,,
\ee
with $I=(0\,,1)$, and $\alpha$ is an index labelling the connected components of $\stsp_{*}$.
It is easy to see that $\stsp_{*}$ is connected, and thus $\alpha=1$.
Indeed, writing the vector $|\psi\rangle$ of which $p_{\psi}$ is the associated pure state as

\be
|\psi\rangle=\sum_{j=1}^{n}\,r_{j}\mathrm{e}^{\imath\vartheta_{j}}\,|j\rangle\,,
\ee
the condition $p_{\psi}\in\stsp_{*}$ implies $r_{j}\neq0$ for all $j$.
From this, it follows that the set of vectors $|\psi\rangle$ such that their associated ray $p_{\psi}$ is in $\stsp_{*}$ is connected.
Then, since the projection $\pi$ from $\mathcal{H}_{0}$ to $P(\mathcal{H})=\stsp$ is continuous, we conclude that $\stsp_{*}$ is connected.

The explicit form for the diffeomorphism $\Psi$ is:

\be
\Psi(p_{\psi})=\left(\mathrm{e}^{\imath(\vartheta_{1}-\vartheta_{n})}\,,\cdots\,,\mathrm{e}^{\imath(\vartheta_{n-1}-\vartheta_{n})}\,,\frac{(r_{1})^{2}}{\sum_{k=1}^{n}(r_{k})^{2}}\,,\cdots\,,\frac{(r_{n-1})^{2}}{\sum_{k=1}^{n}(r_{k})^{2}}\right)\,.
\ee
The $(n-1)$ vector fields $X_{j}$ are tangent to $\mathbb{T}^{n-1}$ and are precisely the fundamental vector fields of the $(n-1)$ tori composing $\mathbb{T}^{n-1}$.
The dynamical vector field is the linear combination:

\be
\Gamma=\sum_{j=1}^{n-1}(\nu_{j} - \nu_{n})X_{j}
\ee
of the canonical vector fields $X_{j}$ with constant coefficients $\nu_{j}$.
Consequently, $\Gamma$ is tangent to $\mathbb{T}^{n-1}$, and the dynamical evolution on $\stsp_{*}$ is the result of $(n-1)$ uncoupled uniform motions on each of the tori. 
If $\nu_{j}\neq\nu_{n}$, the projection $pr^{\mathbb{T}}_{j}$ onto the $j$-th torus provides us with a periodic time function $T_{j}:\stsp_{*}\rightarrow \mathbb{T}$ given by:

\be
T_{j}:=pr^{\mathbb{T}}_{j}\circ\Psi\,,\;\;\;\;\;\;\;\;\;\;T_{j}\circ\phi_{\tau}(p_{\psi})=\mathrm{e}^{\imath\left((\vartheta_{j}-\vartheta_{n})+(\nu_{j} - \nu_{n})\tau\right)}\,.
\ee

Had we started with a different choice of the $(n-1)$ vector fields $X_{j}$ and functions $f_{j}$, we would have got another diffeomorphism:

\be
\Phi:\stsp_{*}\longrightarrow\mathbb{T}^{n-1}\times\mathbb{R}^{n-1}
\ee
and another family of periodic simultaneity functions:

\be
\widetilde{T}_{j}:=pr^{\mathbb{T}}_{j}\circ\Phi\,.
\ee
The relation between $T_{j}$ and $\widetilde{T_{j}}$ can easily be understood.
To see this, let us define the intertwining diffeomorphisms:

\be
\mathcal{I}_{\Psi\Phi}:\stsp_{*}\longrightarrow\stsp_{*}\,,\;\;\;\mathcal{I}_{\Psi\Phi}:=\Phi^{-1}\circ\Psi\,,
\ee
\be
\mathcal{I}_{\Phi\Psi}:\stsp_{*}\longrightarrow\stsp_{*}\,,\;\;\;\mathcal{I}_{\Phi\Psi}:=\Psi^{-1}\circ\Phi\,.
\ee
Clearly, $\mathcal{I}_{\Psi\Phi}^{-1}=\mathcal{I}_{\Phi\Psi}$ and $\mathcal{I}_{\Phi\Psi}^{-1}=\mathcal{I}_{\Psi\Phi}$.
Consequently:

\be
T_{j}=pr^{\mathbb{T}}_{j}\circ\Psi=pr^{\mathbb{T}}_{j}\circ\Psi\circ\mathcal{I}_{\Phi\Psi}\circ\mathcal{I}_{\Psi\Phi}=pr^{\mathbb{T}}_{j}\circ\Phi\circ\mathcal{I}_{\Psi\Phi}=\widetilde{T_{j}}\circ\mathcal{I}_{\Psi\Phi}=\mathcal{I}_{\Psi\Phi}^{*}(\widetilde{T_{j}})\,.
\ee

\vsp

Let us now illustrate the above construction in the case of a $2$-level quantum system, that is, in the case of the Qubit, where $\mathcal{H}\cong\mathbb{C}^{2}$ and the complex projective space is the $2$-dimensional sphere, that is, $P(\mathcal{H})\cong S^{2}$.

Let $\mathbf{H}$ be the Hamiltonian operator of the system, $\nu_{1}$ and $\nu_{2}$ its eigenvalues, and $|1\rangle\,,|2\rangle$ its normalized eigenvectors.
It is clear that the only meaningful dynamical situation corresponds to the case in which $\mathbf{H}$ has a non-degenerate spectrum, otherwise, $\mathbf{H}$ is proportional to the identity operator, its associated dynamical vector field $\Gamma$ on $P(\mathcal{H})$ is the null vector field, and there is no dynamical evolution at all.
Consequently, we assume $\nu_{1}\neq\nu_{2}$.

The pure states $p_{|1\rangle}$ and $p_{|2\rangle}$ corresponding to the normalized eigenvectors of $\mathbf{H}$ are antipodal points on the sphere $P(\mathcal{H})\cong S^{2}$, and they are the only fixed points of the dynamical evolution of the system.
The dynamical trajectories are circles on the sphere with center on the axis passing through $p_{|1\rangle}$ and $p_{|2\rangle}$.
In this case, the reduced space of states $\stsp_{*}$ is the space of states without the fixed points, hence, it has the topology of a cylinder.

If we choose to consider the constant of the motion $e_{\mathbf{E}_{1}}\equiv e_{1}$ associated to $\mathbf{E}_{1}=|1\rangle\langle 1|$, we can write $e_{2}=1-e_{1}$ and:

\be
\Gamma=\left(\nu_{1}-\nu_{2}\right)X_{1}\,,
\ee
with $X_{1}$ the Hamiltonian vector field associated to $e_{1}$.
Then, there is the isomorphism $\Psi:\stsp_{*}\rightarrow \mathbb{T}\times I$:

\be
\Psi(p_{\psi})=\left(\mathrm{e}^{\imath(\vartheta_{1}-\vartheta_{2})}\,,\frac{(r_{1})^{2}}{(r_{1})^{2} + (r_{2})^{2}}\right)\,,
\ee
where $|\psi\rangle=r_{1}\mathrm{e}^{\imath\vartheta_{1}}\,|1\rangle + r_{2}\mathrm{e}^{\imath\vartheta_{2}}\,|2\rangle$, and $r_{1},r_{2}\neq0$.
The periodic time function associated to $\Psi$ reads:

\be
T\circ\phi_{\tau}(p_{\psi})=\mathrm{e}^{\imath\left((\vartheta_{1}-\vartheta_{2})+(\nu_{1} - \nu_{2})\tau\right)}\,.
\ee

On the other hand, if we choose the constant of the motion $e_{\mathbf{E}_{2}}\equiv e_{2}$, we can write

\be
\Gamma=\left(\nu_{2}-\nu_{1}\right)X_{2}\,,
\ee
and we obtain the isomorphism $\Phi:\stsp_{*}\rightarrow \mathbb{T}\times I$:

\be
\Phi(p_{\psi})=\left(\mathrm{e}^{\imath(\vartheta_{2}-\vartheta_{1})}\,,\frac{(r_{2})^{2}}{(r_{1})^{2} + (r_{2})^{2}}\right)\,,
\ee
In this case, the periodic time function associated to $\Phi$ reads:

\be
\widetilde{T}\circ\phi_{\tau}(p_{\psi})=\mathrm{e}^{\imath\left((\vartheta_{2}-\vartheta_{1})+(\nu_{2} - \nu_{1})\tau
\right)}\,.
\ee
In both cases, the sets of simultaneous states are just the meridians on the $2$-dimensional sphere.

\section{Conclusions}

In this contribution, a different approach toward the problem of time in quantum mechanics is proposed.
Motivated by spacetime considerations, we have investigated the possibility of defining a notion of simultaneity in the dynamical context of quantum mechanics.
The main idea is to describe the simultaneity aspect of time for a physical system subject to a dynamical evolution $\{\phi_{\tau}\}$ by specifying all those states that can be interpreted as simultaneous states with respect to the dynamics. 
The essential ingredient in this description is a function $T$ defined on the space of physical states of the system with values in the real numbers $\mathbb{R}$ or in the circle group $\mathbb{T}$, which is equivariant with respect to the dynamical evolution $\{\phi_{\tau}\}$ of the system itself.
The sets of simultaneous states are then defined as the level sets of $T$.
Accordingly, the notion of simultaneity encoded in $T$ is of dynamical nature.

The time function $T$ is introduced in the classical setting, where the space of states $\stsp$ of the system is a finite-dimensional differential manifold, and the dynamical evolution $\{\phi_{\tau}\}$ is the one-parameter group of diffeomorphisms of $\stsp$ generated by a complete vector field $\Gamma$.
Two definitions for a time function $T$ are given, one which is well-suited for dynamical evolutions presenting no periodic orbits, and one which allows to handle periodic orbits.
Some simple examples are then briefly illustrated.

By means of the so-called geometric formulation of quantum mechanics (\cite{ashtekar_schilling-geometrical_formulation_of_quantum_mechanics}, \cite{ercolessi_marmo_morandi-from_the_equations_of_motion_to_the_canonical_commutation_relations}, \cite{grabowski_kus_marmo-geometry_of_quantum_systems_density_states_and_entanglement}, \cite{carinena_ibort_marmo_morandi-geometry_from_dynamics_classical_and_quantum}), the unitary evolutions of pure states of a finite-level quantum system are analyzed.
It is proven that every finite-level quantum system subject to a unitary evolution allows for the definition of a family of periodic time functions.
These functions are intimately connected with the geometrical structure of the system, specifically, to the action-angle variables formulation of the dynamics.

We want to stress two facts concerning these time functions.
First of all, they are not the quantization of some classical time function.
The geometrical definition of the time function given in section \ref{section: Time function and simultaneity for classical systems} applies to classical systems as well as to quantum systems.
Then, they are not the expectation value functions of some operator on the Hilbert space of the system, hence, they are not observables in the canonical sense.
In this regard, the time function introduced here is more similar to the various notions of Entropy, or Purity, or to the various measures of entanglement.

It is worth noting that the canonical commutation relation clearly forbids the existence of a time operator $\mathbf{T}$ for finite-level quantum sytems, while the time function $T$ introduced here always exists for such systems.
This makes $T$ particularly relevant in the context of quantum information theory, where finite-level quantum systems are extensively used.
For instance, an interesting perspective would be to analyze the geometrical properties of the sets of simultaneous states in the case of composite systems, in order to understand if the time function of one of the component systems can be used as a sort of internal clock for monitoring the dynamical evolution of the others in analogy with \cite{page_wootters-evolution_without_evolution_dynamics_described_by_stationary_observables}.
We will consider this situation in the future.

Using the results in \cite{cirelli_lanzavecchia_mania-normal_pure_states_and_the_von_neumann_algebra_of_bounded_operators_as_Kahler_manifold} and \cite{cirelli_mania_pizzocchero-quantum_mechanics_as_an_infinite_dimensional_Hamiltonian_system_with_uncertainty_structure}, the geometrical definition of the time function $T$ could be extended to the case in which the Hilbert space of the system is infinite-dimensional.
This subject is under current investigation.

Finally, the definition of $T$ makes no reference to the infinitesimal generator of the dynamical evolution, hence, a generalization to the case of dissipative dynamics is possible.

\section{Acknowledgements}

Funding: G. M. and J.M. P.P. would like to acknowledge the partial support by the Spanish MINECO grant MTM2014-54692-P and QUITEMAD+, S2013/ICE-2801.

\addcontentsline{toc}{section}{References}
\bibliographystyle{unsrt}
\bibliography{scientific_bibliography}

\end{document}